\documentclass[showpacs,aps,graphicx,twocolumn]{revtex4}
\usepackage{mathrsfs}
\usepackage{amsfonts}
\usepackage{amssymb}
\usepackage{graphicx}
\usepackage{eufrak}
\usepackage{multirow}
\usepackage{amsbsy}

\begin{document}


\title{Synthesis of Multivalued Quantum Logic Circuits by Elementary Gates}

\author{Yao-Min Di$^{1}$\footnote{Corresponding author:
yaomindi@sina.com} and Hai-Rui Wei$^{1, 2}$}
\address{$^{1}$School of Physics $\&$ Electronic Engineering, Jiangsu Normal
University, Xuzhou 221116,  China\\
$^{2}$Department of Physics, Beijing Normal University, Beijing
100875, China }

\date{\today }

\begin{abstract}

We propose the generalized controlled $X$ (\text{\scriptsize{GCX}})
gate as the two-qudit elementary gate, and based on Cartan
decomposition, we also give the one-qudit elementary gates. Then we
discuss the physical implementation of these elementary gates and
show that it is feasible with current technology. With these
elementary gates many important qudit quantum gates can be
synthesized conveniently. We provide  efficient methods for the
synthesis of various kinds of controlled qudit gates and greatly
simplify the synthesis of existing generic multi-valued quantum
circuits. Moreover, we generalize the quantum Shannon decomposition
(QSD), the most powerful technique for the synthesis of generic
qubit circuits, to the qudit case. A comparison of ququart ($d=4$)
circuits and qubit circuits reveals that using ququart circuits may
have an advantage over the qubit circuits in the synthesis of
quantum circuits.

\end{abstract}


\maketitle \flushbottom

\section{Introduction} \label{sec1}

Using multivalued quantum systems (qudits) instead of qubits has a
number of potential advantages. As a specific name, three-level
quantum systems are called qutrits, and four-level systems are also
called ququarts. There have been many proposals to use qudits to
implement quantum computing \cite{1,2,3,4,5,6}. Now there is an
increasing interest in this area, and some experimental works on
qudit systems have been developed in recent years \cite{7,8,9}.

Many works have been done in multivalued quantum logic synthesis.
Brylinski and Brylinski \cite{10} and Bremner \emph{et al}.
\cite{11} concluded that any two-qudit gate that creates
entanglement without ancillas can act as a universal gate for
quantum computation, when assisted by arbitrary one-qudit gates.
Brennen \emph{et al}. proposed  use of the controlled increment
(\text{\scriptsize{CINC}}) gate as a two-qudit elementary gate,
investigated the synthesis of general qudit circuits based on
spectral decomposition, and the ``Triangle'' algorithm \cite{4,5},
and obtained asymptotically optimal results, but for the synthesis
of specific qudit gates, using this gate is inconvenient and the
relevant work is rarely seen. There are other proposals, such as the
\text{\scriptsize{GXOR}} \cite{3}, \text{\scriptsize{SUM}} \cite{6},
etc., but no practice circuits are given. The synthesis of binary
quantum circuits has been extensively investigated by many authors
\cite{12,13,14,15,16,17,18,19,20,21,22}, and it is rather mature
now. In the previous work for qudit circuits the methods in qubit
circuits are seldom used. Since there are technical difficulties
\cite{23} with the tensor product structure of qudits, whether these
methods are useful for qudits has been an open question. Moreover,
there is no unified measure for the complexity of qubit and various
qudit circuits yet, which makes it inconvenient to compare them.

In this article we focus on the synthesis of multivalued quantum
logic circuits. With the elementary gates proposed here we can
synthesize many specific qudit quantum gates conveniently, greatly
simplify the synthesis of existing generic multi-valued quantum
circuits, generalize the quantum Shannon decomposition (QSD)
\cite{20}, the most powerful technique for the synthesis of generic
qubit circuits, to the qudit case and get many best known results.
Moreover, the defects mentioned above are all overcome.

The article is organized as follows. In Sec. \ref{sec2} we propose
the general controlled $X$ (\text{\scriptsize{GCX}}) gate as a
two-qudit elementary gate, and based on Cartan decomposition
\cite{24} we also give a set of one-qudit elementary gates. They can
be used as a unified measure of complexity for various quantum logic
circuits. In Sec. \ref{sec3} we investigate the physical
implementation of these gates and show that it is feasible with
current technology. With these gates we investigate the synthesis of
some important multivalued quantum gates and the synthesis of
various controlled qudit gates in Sec. \ref{sec4}. We generalize the
QSD to qudit case in Sec. \ref{sec5}, revealing that using ququart
circuits may have an advantage over the qubit circuits in the
synthesis of quantum circuits. Finally, a brief conclusion is given
in Sec. \ref{sec6}. The Cartan decomposition used in Sec. \ref{sec2}
is given in Appendix \ref{A}.

\section{Elementary gates} \label{sec2}

\begin{figure}[!h]
\begin{center}
\includegraphics[width=2.5 cm,angle=0]{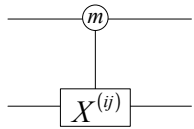}
\caption{Generalized controlled-$X$ gate.}              \label{Fig1}    
\end{center}
\end{figure}

\begin{center}

\begin{figure}[!h]
\begin{center}
\includegraphics[width=6.5 cm,angle=0]{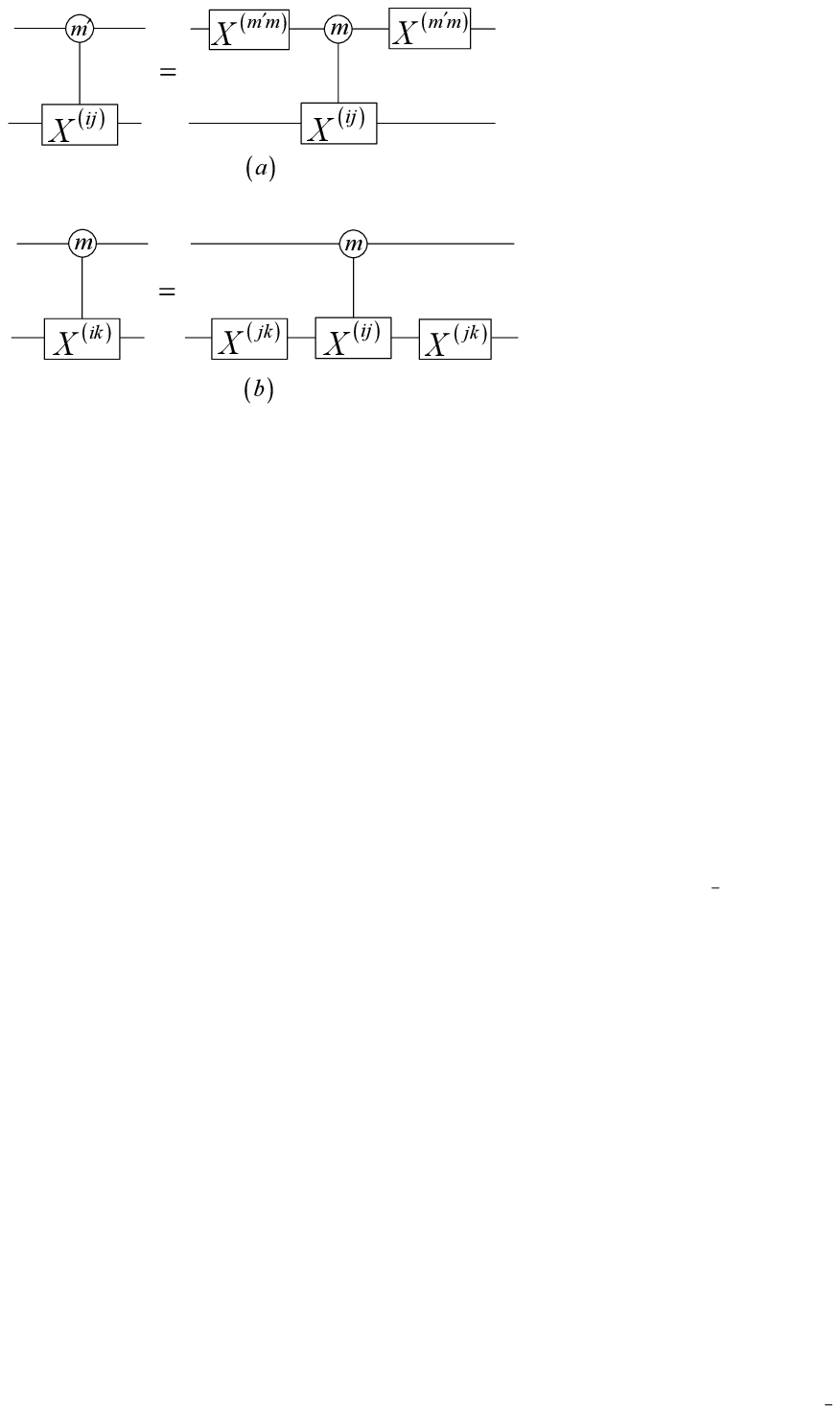}
\caption{Transformation among different \text{\scriptsize{GCX}}
gates:
(a) transformation of control mode. (b) transformation of target operations.}    \label{Fig2}          
\end{center}
\end{figure}
\end{center}

There are $d(d-1)/2$ single-qudit $X^{(ij)}$ gates which act on the
two-dimensional subspace $\mathscr{H}_{ij}$ of $d$-dimensional
Hilbert space, where $X^{(ij)}=|i\rangle\langle j|+|j\rangle\langle
i|+\sum_{k\neq i,j}|k\rangle\langle k|$. The \text{\scriptsize{GCX}}
gate is the two-qudit gate which implements the $X^{(ij)}$ operation
on the target qudit iff the control qudit is in the state
$|m\rangle$, $(m\in\{0,1,\ldots,d-1\})$. The circuit representation
for the \text{\scriptsize{GCX}} gate is shown in Fig. \ref{Fig1}, in
which the line with a circle represents the control qudit while that
with a square the target qudit. There are $d^2(d-1)/2$ different
forms of the gate and they can be easily transferred to one another
as shown in Fig. \ref{Fig2}.


The \text{\scriptsize{CINC}} gate is a controlled one-qudit gate
which implements the \text{\scriptsize{INC}} operation on the target
qudit iff the control qudit is in the states $|m\rangle$, where
INC$|j\rangle=|j+1,mod\; d\rangle$.  The \text{\scriptsize{INC}}
operation can be decomposed into $d-1$ $X$ operations, so the
\text{\scriptsize{CINC}} gate can be synthesized by $d-1$
\text{\scriptsize{GCX}} gates. The \text{\scriptsize{GCX}} gate is
an elementary counterpart of the binary \text{\scriptsize{CNOT}}
gate, so we propose the \text{\scriptsize{GCX}} gate as the
two-qudit elementary gate for multivalued quantum computing. It can
be used as a unified measure for the complexity of various quantum
circuits.

Suppose $M$ is the matrix of a one-qutrit gate. Take a kind of AIII
type Cartan decomposition \cite{23} of the $\mathrm{U}(3)$ group,
which can be expressed as
\begin{eqnarray}      \label{eq1}
M=e^{i\varphi}M_1^{(jk)}M^{(j'k')}M_2^{(jk)}.
\end{eqnarray}
Here $M^{(jk)}$ is a special unitary transformation in
two-dimensional subspace $\mathscr{H}_{jk}$, and it can be factored
further by the Euler decomposition. The Euler decomposition usually
has two modes: $ZYZ$ decomposition and $XYX$ decomposition. So the
set of one-qutrit elementary gates has two pairs of basic gates,
$R_y^{(jk)}$, $R_z^{(jk)}$, $R_y^{(j'k')}$, $R_z^{(j'k')}$ or
$R_x^{(jk)}$, $R_y^{(jk)}$, $R_x^{(j'k')}$, $R_y^{(j'k')}$. Here
$R_\alpha^{(jk)}(\theta)=\exp(-i\theta\sigma_\alpha^{(jk)}/2)$, for
$0\leq j<k\leq2$, $\alpha\in \{x,y,z\}$, and
$\sigma_x^{(jk)}=|j\rangle\langle k|+|k\rangle\langle j|$,
$\sigma_y^{(jk)}=-i|j\rangle\langle k|+i|k\rangle\langle j|$,
$\sigma_z^{(jk)}=|j\rangle\langle j|-|k\rangle\langle k|$.

Using successive AIII-type Cartan decompositions of the
$\mathrm{U}(d)$ group, a generic one-qudit gate can be decomposed to
a series of $M^{(jk)}$, which involves at least $d-1$ kinds of
$M^{(jk)}$ that act on different 2D subspaces. To implement a qudit
gate requires $d-1$ driving fields, and $M^{(jk)}$s essentially are
single-qubit gates. So the set of one-qudit elementary gates has
$d-1$ pairs of $R_\alpha^{(jk)}$ gates acting on $d-1$ different 2D
subspaces. The choice of $d-1$ pairs of basic gates is not unique.
They are universal if only the corresponding driving fields can
connect the $d$ levels of the qudit together.

\section{Physical implementation} \label{sec3}

In the last decade, there has been tremendous progress in the
experimental development of qubit quantum computing, and the problem
of constructing a \text{\scriptsize{CNOT}} gate has been addressed
from various perspectives and for different physical systems
\cite{25,26,27,28,29,30,31,32,33,34,35,36}. The
\text{\scriptsize{GCX}} gate is essentially binary, so it can be
implemented with existing technique.

\begin{figure}[!h]
\begin{center}
\includegraphics[width=3.5 cm,angle=0]{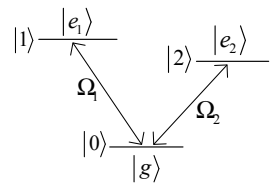}
\caption{V-type three level quantum system.}   \label{Fig3}
\end{center}
\end{figure}

Assume we have a V-type three-level quantum system shown in Fig.
\ref{Fig3}, which constitutes a qutrit and the two levels of the
system $|0\rangle$ and $|1\rangle$ forms a qubit. Two laser beams
$\Omega_{1}$ and $\Omega_{2}$  are applied to the ion to manipulate
$|0\rangle\leftrightarrow|1\rangle$
  and $|0\rangle\leftrightarrow|2\rangle$ transition, respectively. If a two-qubit CNOT gate is realized
in such systems, one \text{\scriptsize{GCX}} gate is naturally
obtained, and the eight other \text{\scriptsize{GCX}} gates formed
can be obtained by the transformation shown in Fig. \ref{Fig2}. The
single-qutrit gates are implemented by Rabi oscillations between the
qutrit levels. Applying the laser pulses in $\Omega_{1}$ and
$\Omega_{2}$ and choosing suitable phases, this allows us to perform
$R_{x}^{(01)}$, $R_{y}^{(01)}$ and $R_{x}^{(02)}$, $R_{y}^{(02)}$
gates, respectively \cite{37,38}. So a set of one-qutrit elementary
gates is obtained, and any one-qutrit gate can be implemented
according to Eq. (\ref{eq4}). There are  two other types of quantum
system: the  $\Lambda$ type and cascade type. We can use
$R_{x}^{(01)}$, $R_{y}^{(01)}$, $R_{x}^{(12)}$, $R_{y}^{(12)}$ or
$R_{y}^{(02)}$, $R_{z}^{(02)}$, $R_{y}^{(12)}$, $R_{z}^{(12)}$ as
one-qutrit elementary gates to meet the requirement of manipulating
quantum states in these types of quantum systems. The method can be
naturally generalized to the generic qudit case.

It is not too difficult to find such a quantum system. Early in
2003, the Innsbruck group implemented the complete Cirac-Zoller
protocol \cite{25} of the \text{\scriptsize{CONT}} gate with two
calcium ions ($\mathrm{Ca^{+}}$) in a trap \cite{27}.  The original
qubit information is encoded in the ground-state $S_{1/2}$ and
metastable $D_{5/2}$ state. The $D_{5/2}$ state has a lifetime $\tau
\backsimeq 1.16$ s. There is another metastable $D_{3/2}$ state in
$\mathrm{Ca^{+}}$. Its lifetime is about the same as that of the
$D_{5/2}$ state. The three levels of $\mathrm{Ca^{+}}$, one ground
state and two metastable states, may constitute a qutrit candidate.
The CNOT gate was implemented by Schmidt-Kaler \emph{et al.}
\cite{27} and forms naturally a \text{\scriptsize{TCX}} gate. Two
laser pulses are used to manipulate the $S_{1/2}\leftrightarrow
D_{5/2}$ quadruple transition near 729 nm and the
$S_{1/2}\leftrightarrow D_{3/2}$ transition near 732 nm,
respectively. Rabi oscillations between these levels can implement
the one-qutrit elementary gates $R_x^{(01)}$, $R_y^{(01)}$ and
$R_x^{(02)}$, $R_y^{(02)}$.

The superconducting quantum information processing devices are
typically operated as qubit by restricting them to the two lowest
energy eigenstates. By relaxing this restriction, we can operate it
as a qutrit or qudit. The experimental demonstrations of the
tomography of a transmon-type superconducting qutrit have been
reported in \cite{9}, and the emulation of a quantum spin greater
than $1/2$ has been implemented in a superconducting phase qudit
\cite{8}. This means that to prepare a one-qutrit state or one-qudit
state and a read out on these systems has been implemented, so the
one-qutrit gates or one-qudit gates can also be implemented on the
systems. Construction of a robust \text{\scriptsize{CNOT}} gate on
superconducting qubits has been extensively investigated
\cite{34,35,36}, which means that the condition to implement
multivalued quantum computing has come to maturity on these
superconducting devices.

\section{Synthesis of multi-valued quantum logic gates}   \label{sec4}

\subsection{Synthesis of some important multi-valued quantum gates}   \label{sec41}

By using \text{\scriptsize{GCX}} gates, some important qudit gates
can be synthesized conveniently. The reason is that the
$X^{(i,\,i+1)}$'s operations are the generators of the permutation
group $S_d$, while \text{\scriptsize{INC}}, etc. operations are not.
The multivalued \text{\scriptsize{SWAP}} gate interchanges the
states of two qudits acted on by the gate. The ternary
\text{\scriptsize{SWAP}} gate can be decomposed into three binary
\text{\scriptsize{SWAP}} gates, that is
\begin{eqnarray}    \label{eq2}
W=W^{(01)}\cdot W^{(02)}\cdot W^{(12)}.
\end{eqnarray}
Here $W^{(ij)}=|ij\rangle\langle ji|+|ji\rangle\langle
ij|+\sum_{kl\neq ij,ji}|kl\rangle\langle kl|$, and it can be
synthesized by three \text{\scriptsize{GCX}} gates. So the ternary
\text{\scriptsize{SWAP}} gate is synthesized by nine
\text{\scriptsize{GCX}} gates. For the generic qudit case, the
multivalued \text{\scriptsize{SWAP}} gate can be decomposed into
$d(d-1)/2$  binary \text{\scriptsize{SWAP}} gates, each of them
needing three \text{\scriptsize{GCX}} gates. The multivalued root
\text{\scriptsize{SWAP}} gate can also be decomposed into $d(d-1)/2$
binary root \text{\scriptsize{SWAP}} gates.

\begin{figure}[!h]
\begin{center}
\includegraphics[width=5.5 cm,angle=0]{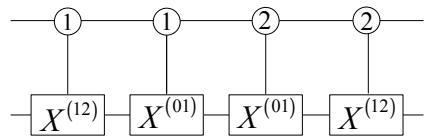}
\caption{Synthesis of ternary \text{\scriptsize{SUM}} gate.}
\label{Fig4}
\end{center}
\end{figure}

\begin{figure}[!h]
\begin{center}
\includegraphics[width=4.5 cm,angle=0]{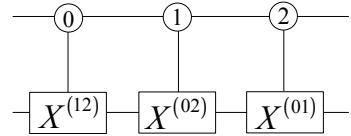}
\caption{Synthesis of ternary \text{\scriptsize{GXOR}} gate.}
\label{Fig5}
\end{center}
\end{figure}

We denote two inputs of a $d$-dimensional two-qudit as $A$ and $B$,
respectively. The \text{\scriptsize{SUM}} gate is a two-qudit gate
in which an output remains $A$ unchanged, and another output is the
sum of $A$ and $B$ modulo $d$ denoted $A\oplus B$. The
\text{\scriptsize{GXOR}} gate is similar to the
\text{\scriptsize{SUM}} gate. The difference is that the output is
the difference of $A$ and $B$ modulo $d$. The synthesis of the
ternary \text{\scriptsize{SUM}} gate and ternary
\text{\scriptsize{GXOR}} gate base on the \text{\scriptsize{GCX}}
gates is shown in Fig. \ref{Fig4} and Fig. \ref{Fig5}, respectively.

\begin{widetext}
\begin{figure}[!h]
\begin{center}
\includegraphics[width=17.50 cm,angle=0]{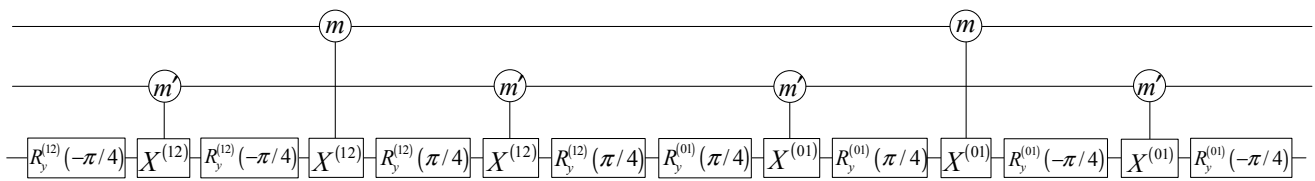}
\caption{Synthesis of the ternary
$\wedge_2$(\text{\scriptsize{INC}}) gate.}\label{Fig6}
\end{center}
\end{figure}
\end{widetext}

The twofold generalized controlled $X$ gate [$\wedge_2(X)$] is a
three-qudit gate in which two control qudits are unaffected by the
action of the gate, and the target qudit is acted on by the
$X^{(ij)}$ operation iff the two control qudits are in the states
$|m\rangle$, $|m'\rangle$ respectively. It is essentially a Toffoli
gate \cite{37}, which can be synthesized with six
\text{\scriptsize{GCX}} gates and ten single-qudit gates acting on a
2D subspace. In some cases, we can use the psuedo-$\wedge_2(X)$ gate
$\left(p\wedge_2\left(X\right)\right)$ instead of the $\wedge_2(X)$
gate. The $p\wedge_2(X)$ gate is also a three-qudit gate that two
control qudits are unaffected by the action of the
 gate, the target qudit is acted by the $X^{(ij)}$ operation iff the
two control qudits are in the states $|m\rangle$,  $|m'\rangle$
respectively and by the $Z^{(ij)}$ or $Z^{(ji)}$ operation iff the
first control qudit is in the state $|m\rangle$, the second control
qudit is not in the state $|m'\rangle$. It is synthesized by three
\text{\scriptsize{GCX}} gates and two $R_\alpha^{(ij)}(\pi/4)$  and
two $R_\alpha^{(ij)}(-\pi/4)$ gates  (see Appendix \ref{B}).  The
two-fold controlled \text{\scriptsize{INC}} gate
[$\wedge_2$(\text{\scriptsize{INC}})] is that the two qudits remain
no change, the qudit is acted by the \text{\scriptsize{INC}}
operation iff two control qudits are in the control states
$|m\rangle$, $|m'\rangle$ respectively. The ternary
$\wedge_2$(\text{\scriptsize{INC}}) gate consists of two
$p\wedge_2(X)$ gates, and the synthesis is shown in Fig. \ref{Fig6},
which requires six \text{\scriptsize{GCX}} gates and eight
$R_y^{(ij)}(\theta)$ gates. In  $d$-valued qudit case, the synthesis
of $\wedge_2$(\text{\scriptsize{INC}}) gate requires $3(d-1)$
\text{\scriptsize{GCX}} gates for $d$ is odd, and $3d$
\text{\scriptsize{GCX}} gates for $d$ is even. It is much simpler
than that in \cite{5}. That needs $(d+2)d$ \text{\scriptsize{CINC}}
gates and $(d+1)d$ \text{\scriptsize{CINC}}$^{-1}$ gates, which is
equivalent to $(2d+3)d(d-1)$ \text{\scriptsize{GCX}} gates.

\subsection{Synthesis of various controlled qudit gates}   \label{sec42}

\begin{figure}[!h]
\begin{center}
\includegraphics[width=5.0 cm,angle=0]{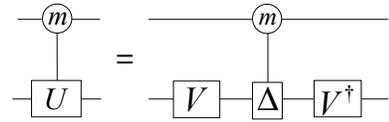}
\caption{Synthesis of a controlled $U$ gate.}      \label{Fig7}
\end{center}
\end{figure}

\begin{widetext}
\begin{figure}[!h]
\begin{center}
\includegraphics[width=15.0 cm,angle=0]{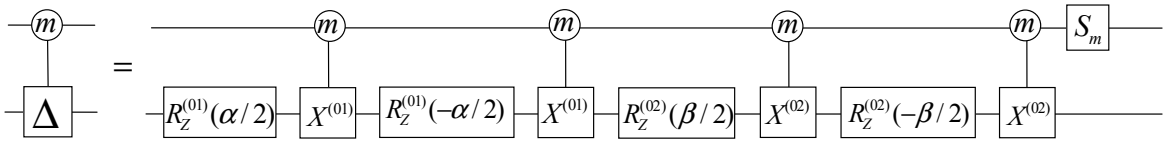}
\caption{Synthesis of a controlled diagonal qutrit gate.}
\label{Fig8}
\end{center}
\end{figure}
\end{widetext}

A controlled one-qudit gate [$\wedge_1(U)$] is a two-qudit gate in
that iff the control qudit is set to the state $|m\rangle$ then  a
unitary operation $U$ is applied to the target qudit. From the
diagonal decomposition $U=VDV^\dagger$, we can get a synthesis of a
controlled $U$ gate which involves a pair of one-qudit gates and a
controlled diagonal [$\wedge_1(\bigtriangleup)$] gate as shown in
Fig.\ref{Fig7}. Here $V$ is unitary, and $D$ is diagonal and has the
form
\begin{eqnarray}            \label{eq3}
D&=&e^{\varphi}diag\{e^{-i(\alpha_1+\alpha_2+\ldots+\alpha_{d-1})},e^{i\alpha_1},e^{i\alpha_2},\ldots,e^{i\alpha_{d-1}}\}\nonumber\\
&=&e^{\varphi}R_z^{(01)}(\alpha_1)R_z^{(02)}(\alpha_2)\ldots R_z^{(0
(d-1))}(\alpha_{d-1}).
\end{eqnarray}
The $\wedge_1(\bigtriangleup)$ gate can be synthesized by a phase
qudit and $d-1$ controlled $R_z^{(ij)}$ gates, each of them needing
two \text{\scriptsize{GCX}} gates. Hence the generic $\wedge_1(U)$
 gate can be synthesized by $2(d-1)$ \text{\scriptsize{GCX}} gates
in the worst case. In qutrit case the synthesis of a
$\wedge_1(\bigtriangleup)$ gate is shown in Fig. \ref{Fig8}, where
$S_m=\sum_j(1+\delta_{jm}(e^{i\varphi}-1)|j\rangle\langle j|$ is a
phase qutrit gate.

\begin{figure}[!h]
\begin{center}
\includegraphics[width=8.0 cm,angle=0]{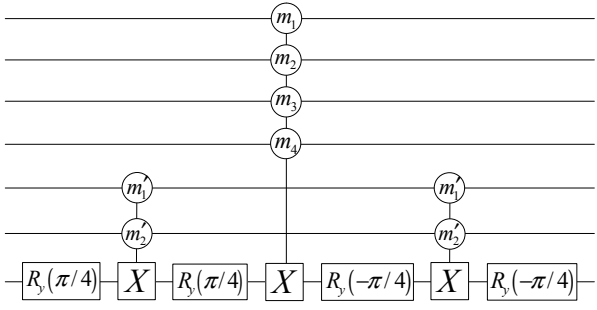}
\caption{A circuit implementing a $p\wedge_k(X)$ gate.} \label{Fig9}
\end{center}
\end{figure}

\begin{figure}[!h]
\begin{center}
\includegraphics[width=4.0 cm,angle=0]{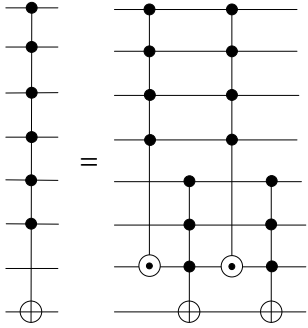}
\caption{A circuit implementing $\wedge_k(X)$ gate. For brief, the
symbols in the qubit circuit are used and the $\odot$ denotes the
pseudo-$X$ operation.}   \label{Fig10}
\end{center}
\end{figure}

The $k$-fold controlled one-qudit gate [$\wedge_k(U)$] has $k$
control qudits and a target qudit. Similar to the synthesis of the
$\wedge_1(U)$ gate, a $\wedge_k(U)$ gate is composed of a pair of
one-qudit gates and a $k$-fold controlled one-qudit diagonal
one-qudit gate $[\wedge_k(\bigtriangleup)]$. The
$\wedge_k(\bigtriangleup)$ gate can be synthesized by a $(k-1)$-fold
controlled phase qudit and $d-1$ $\wedge_k(R_z)$ gates, and each
$\wedge_k(R_z)$ needs a pair of $\wedge_k(X)$ gates. To simplify the
synthesis of $\wedge_k(U)$ gates, we introduce the
pseudo-$\wedge_k(X)$ [$p\wedge_k(X)$] gates. The $p\wedge_k(X)$ gate
has two sets of control qudits. Its target qudit is acted by the
$X^{(ij)}$ operation iff the two sets of control qudits are in the
control states $|m_1,m_2,\cdots,m_{k_{1}}\rangle$ and
$|m_1^{\prime}, m_2^{\prime},\cdots,  m_{k_{2}}^{\prime}\rangle$,
respectively, and by the $Z^{(ij)}$ operation iff the first set of
control qudits is in the control states and the second set of
control qudits is not in the control states, where $k_1+k_2=k$. Now
we present a scheme for implementing $p\wedge_k(X)$ gates and a
scheme for implementing $\wedge_k(X)$ gates, shown in Fig.
\ref{Fig9} and
 \ref{Fig10}, respectively. Since the $\wedge_k(X)$  gates appear in
 the $\wedge_k(R_z)$ gate in a pair and the $R_z$ gates are diagonal,
they can be replaced by $p\wedge_k(X)$ gates. The $k$-fold
controlled unimodular one-qudit gate can be synthesized by
$2p_k(d-1)$ \text{\scriptsize{GCX}} gates. Here $p_k$ denotes the
numbers of \text{\scriptsize{GCX}} gates in a $p\wedge_k(X)$ gate,
and it can be obtained by its recursive implementing process. The
$(k-1)$-fold controlled phase gate can be further decomposed into a
$(k-1)$-fold controlled unimodular diagonal gate and a $(k-2)$-fold
controlled phase gate. By successive decomposition we can get that
the synthesis of a $k$-fold controlled general one-qudit gate
requires $2(1+\sum_{s=2}^{k} p_s)(d-1)\leq (1+\delta_{n2})k^3 (d-1)$
\text{\scriptsize{GCX}} gates. The estimate
$2(1+\sum_{s=2}^{k}p_s)\leq(1+\delta_{n2})k^3$ comes from practice
data.

With efficient synthesis of $\wedge_1(U)$ and $\wedge_k(U)$ gates we
can greatly simplify the synthesis of existing multiqudit circuits.
Based on the spectral decomposition, for the circuit without
ancillas, the \text{\scriptsize{GCX}} count of generic $n$-qudit
circuits is
\begin{eqnarray}                \label{eq4}
C_s&=&2d^n
[(d^n-1)-n(d-1)]+2(1+\sum_{s=2}^{n-1}p_s)d^n(d-1)\nonumber\\
&\leq&2d^n[(d^n-1)-n(d-1)]\nonumber\\&+&(1+\delta_{n2})(n-1)^3d^n(d-1),
\end{eqnarray} whereas the \text{\scriptsize{CINC}} count using the
spectral decomposition given in \cite{5} is
\begin{eqnarray}            \label{eq5}
\ell_s\leq2d^{n+1}[(d^n-1)/(d-1)-n]+(n+1)^{2+\log_2d}d^{n+4}.
\end{eqnarray}

\section{Quantum shannon decomposition} \label{sec5}

\begin{figure}[!h]
\begin{center}
\includegraphics[width=8.5 cm,angle=0]{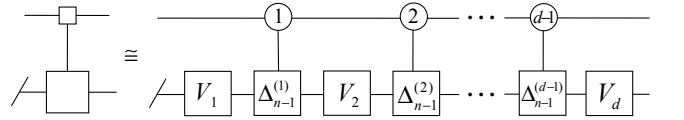}
\caption{Decomposition of a uniformly controlled multiqudit gate.
Here the small square ($\Box$) denotes uniform control and the slash
(/) represents multiple qudits on the line.} \label{Fig11}
\end{center}
\end{figure}

 A $n$-qudit gate corresponds to a $d^n \times d^n$ unitary
matrix. Using Cosine-sine decomposition (CSD) \cite{14,39} we
decompose it to $d^{n-1}\times d^{n-1}$ block diagonal matrices and
cosine-sine matrices. The block diagonal matrix is a uniformly
controlled multi-qudit gate, which can be reduced to a $(n-1)$-qudit
gate and $d-1$ copies of controlled  ($n-1$)-qudit
$[\wedge_1(U(d^{n-1}))]$ gates. It can be further reduced to $d$
copies of $(n-1)$-qudit gates and $d-1$ copies of ($n-1$)-qudit
diagonal $(\wedge_1(\Delta_{n-1}))$ gates as shown in Fig.
\ref{Fig11}. Taking $d=2$, the related decomposition of a block
diagonal matrix is
\begin{eqnarray}             \label{eq6}
\label{1} \left(\begin{array}{cc}
U_1&\\
&U_2\\
\end{array}\right)=
\left(\begin{array}{cc}
V_1&\\
&V_1\\
\end{array}\right)
\left(\begin{array}{cc}
I&\\
&\Delta_{n-1}\\
\end{array}\right)
\left(\begin{array}{cc}
V_2&\\
&V_2\\
\end{array}\right).
\end{eqnarray}
It is equivalent to the decomposition
\begin{eqnarray}               \label{eq7}
\label{1} \left(\begin{array}{cc}
U_1&\\
&U_2\\
\end{array}\right)=
\left(\begin{array}{cc}
W&\\
&W\\
\end{array}\right)
\left(\begin{array}{cc}
D&\\
&D^{\dag}\\
\end{array}\right)
\left(\begin{array}{cc}
V&\\
&V\\
\end{array}\right),
\end{eqnarray}
where  $W=V_1\Delta_{n-1}^{1/2}$, $D=\Delta_{n-1}^{-1/2}$ and
$V=V_2$. It is just  the decomposition of block diagonal matrices in
QSD. So the decomposition given here for qudit circuit can be
considered as a generalized QSD.

\begin{figure*}[!htb]
\begin{center}
\includegraphics[width=17.5 cm,angle=0]{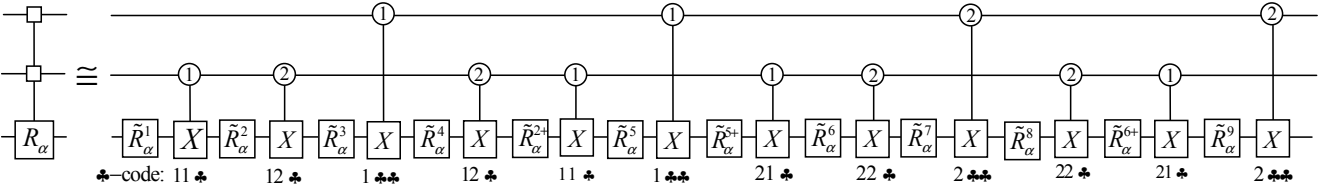}
\caption{Quantum circuit implementing ternary uniformly two-fold
controlled  $R_\alpha^{(ij)}(\alpha\in y, z)$ rotation. Here we use
$\tilde{R}_\alpha^k=R_\alpha^{(ij)}(\theta_k)$, and the $\clubsuit$
codes without zero entry are used to define the position of the
control nodes of \text{\scriptsize{GCX}} gates.}   \label{Fig12}
\end{center}
\end{figure*}

\begin{figure}[!htb]
\begin{center}
\includegraphics[width=8.5 cm,angle=0]{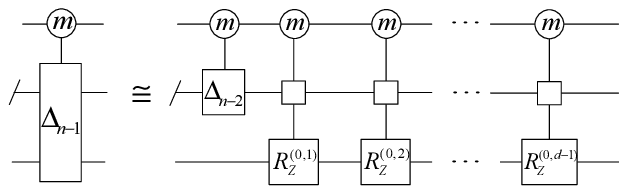}
\caption{Quantum circuit implementing a
$\wedge_1(\bigtriangleup_{n-1})$ gate}     \label{Fig13}
\end{center}
\end{figure}

Here we give a very efficient synthesis of the multivalued uniformly
multifold controlled $R_\alpha^{(ij)} (\alpha \in y, z)$ rotation.
The method parallels the techniques in \cite{14,40} for the qubit
case. For $d=3$ and $n=3$, its synthesis is shown in Fig.
\ref{Fig12}, and in the generic case, it needs $2d^{n-2}(d-1)$
\text{\scriptsize{GCX}} gates (see Appendix \ref{C}). The circuit
can be conveniently obtained by the $\clubsuit$ sequence \cite{4}.
To divide the elements of a $\wedge_1(\bigtriangleup_{n-1})$ gate
into $d^{n-2}$ groups and factor out a phase for each group to make
it unimodular, we get a circuit of the
 gate as shown in Fig.\ref{Fig13}.
It can be further inferred that it needs $2(d^{n-1}-1)$
\text{\scriptsize{GCX}} gates for the synthesis of a
$\wedge_1(\bigtriangleup_{n-1})$ gate.

\begin{table*}[!htbp]
\caption {Exact \text{\scriptsize{GCX}} gate count for the synthesis
of qudit quantum circuits obtained using the QSD decomposition. Here
$\divideontimes$ means this count is better than that using the
spectral algorithm obtained by Eq.(\ref{eq4}).\\}
\begin{ruledtabular}
\begin{tabular}{cccccccc}

 $$      &     $d=3$       &   $d=4$                            &   $ d= 5$                            &   $d=6 $                            &  $d=7$                             & $d=8$ \\\hline

2 & $\divideontimes$44      & $\divideontimes$108                & $\divideontimes$272                  & $\divideontimes$510                 & $\divideontimes$828                & $\divideontimes$1176     \\

3 & $\divideontimes$692     & $\divideontimes$2232               & $\divideontimes$10256                & $\divideontimes$25860               & $\divideontimes$52740              & $\divideontimes$85456   \\

4 &$\divideontimes$6860     & $\divideontimes$37800              & $\divideontimes$336144               & $\divideontimes$1158720             & $\divideontimes$2965788            & $\divideontimes$5551504 \\

5 &$\divideontimes$83924    & $\divideontimes$613248             & $\divideontimes$10796560             & $\divideontimes$51109320            & $\divideontimes$166400964          & $\divideontimes$355955600     \\

6 &$\divideontimes$1011932  & $\divideontimes$7392768            & $\divideontimes$345689872            & $\divideontimes2.25\times10^{9}$    & $\divideontimes9.32\times10^{9}$   & $\divideontimes2.28\times10^{10}$    \\

7 & 12157748                & $\divideontimes$118419456          & $\divideontimes1.11\times10^{10}$    & $\divideontimes9.90\times10^{10}$   & $\divideontimes5.22\times10^{11}$   & $\divideontimes1.46\times10^{12}$    \\

8 &145936700                & $\divideontimes 1.90\times10^{9}$  & $3.55\times10^{11}$                  & $\divideontimes4.36\times10^{12}$   & $\divideontimes2.92\times10^{13}$    & $\divideontimes9.34\times10^{13}$    \\
\end{tabular}               \label{Table1}
\end{ruledtabular}
\end{table*}

\begin{table*}[!htbp]
\caption {A Comparison of gate counts of ququart circuits and qubit
circuits \cite{20} based on QSD. The index $l$ denotes the recursion
bottoms, with which the results come out, at one-qubit or
one-ququart gates ($l = 1$), or two-qubit gates ($l = 2$).\\}
\begin{ruledtabular}
\begin{tabular}{ccccccc}
                                   & 1     &   2     &   3       &   4         &  n                  \\\hline

$n$-ququart gate  ($l=1$)          & 0     & 108     & 2232      &37800        &  $(47/80)\times 4^{2n}-(11/4)\times 4^n +8/5 $   \\

$2n$-qubit gate  ($l=1$)           & 6     & 168     & 2976      & 48768       &  $(3/4)\times 4^{2n}-(3/2)\times 4^n $ \\

$2n$-qubit gate  ($l=2$)           & 3     & 120     & 2208      & 36480       &  $(9/16)\times 4^{2n}-(3/2)\times 4^n$ \\

$2n$-qubit gate  ($l=2$, optimal)  & 3     & 100     & 1868       & 30927      &  $(23/48)\times 4^{2n}-(3/2)\times 4^n +4 /3 $ \\
\end{tabular}               \label{Table2}
\end{ruledtabular}
\end{table*}

Taking $d=4$ as an example, using CSD, the matrix of an $n$-ququart
circuit can be decomposed into four block diagonal matrices and
three cosine-sine matrices. Each of the block diagonal matrices
involves four ($n-1$)-ququart gates and three controlled diagonal
($n-1$)-ququart gates, and each of the cosine-sine matrices involves
two uniformly ($n-1$)-fold controlled $R_y$ rotations. So a generic
$n$-ququart circuit involves 16 ($n-1$)-ququart gates, 12 controlled
diagonal ($n-1$)-ququart gates, and six uniformly ($n-1$)--fold
controlled $R_y$ rotations. From these, we can calculate the
\text{\scriptsize{GCX}} gate count based on QSD.

The exact \text{\scriptsize{GCX}} counts based on generalized QSD
are tabulated in Table \ref{Table1}. When the number of qudits $n$
is small, it gives the simplest known quantum circuit, and when $d$
is a power of two, the circuits given here have the best known
asymptotic features. The $n$-ququart ($d=4$) gate is needed
asymptotically $O(47/80\times4^{2n})$ \text{\scriptsize{GCX}} gates,
whereas it needs asymptotically $O(2\times4^{2n})$
\text{\scriptsize{GCX}} gates based on a spectral algorithm.
Moreover, we compare ququart circuits with qubit circuits based on
QSD in Table \ref{Table2}. Here the gate counts of a generic
$n$-ququart circuit are obtained with recursion bottoms out at
one-ququart gates ($l=1$); they are smaller than that of the
corresponding $2n$-qubit circuit ($l=1)$. The counts can be improved
further by finding more efficient synthesis of two-qudit gates and
using some special optimal techniques. From this, the advantage of
ququart over qubit in the synthesis of generic quantum circuits has
been presented in the first time.

\section{CONCLUSION} \label{sec6}

We propose the \text{\scriptsize{GCX}} gate as the twoqudit
elementary gate of multivalued quantum circuits, and based on Cartan
decomposition, the one-qudit elementary gates are also given. They
are simple, efficient, and easy to implement. With these gates,
various qudit circuits can be efficiently synthesized. Moreover, it
can be used as a unified measure for the complexity of various
quantum circuits. So the crucial issue of which gate is chosen as
the elementary gate of qudit circuits is addressed. In spite of the
difficulties with the tensor product structure of qudits, the
methods used in qubit circuits still can play a very important role.
The comparison of ququart circuits and qubit circuits based on QSD
reveals that using ququart circuits may have an advantage over the
qubit circuits in the synthesis of quantum circuits.

Multivalued quantum computing is a new and exciting research area.
In the synthesis of multivalued quantum circuits there is still
plenty of work to do. It will further reveal the advantage of qudit
circuits over the conventional qubit circuits. Choosing a suitable
quantum system, such as trapped ions, superconducting qudits, and
quantum dots, to investigate the physical implementation of
multivalued quantum logic gates and  undertaking the experimental
work is crucial for the development of multivalued quantum
information science.

\section*{ACKNOWLEDGEMENTS}

This work is supported by the National Natural Science Foundation of
China under Grant No. 61078035 and the Priority Academic Program for
the Development of Jiangsu Higher Education Institutions.

\appendix

\section{CARTAN DECOMPOSITON} \label{A}

The Cartan decomposition of a Lie group depends on the decomposition
of its Lie algebras \cite{24}. Let $\mathfrak{g}$ be a semisimple
Lie algebra and there is the decomposition
\setcounter{equation}{0}
\begin{equation}                                    \label{A1}
\mathfrak{g}=\mathfrak{l}\oplus \mathfrak{p},
\end{equation}
where $\mathfrak{l}$ and $\mathfrak{p}$ satisfy the commutation
relations
\begin{eqnarray}                                                        \label{A2}
[\mathfrak{l},\mathfrak{l}]\subseteq \mathfrak{l},
[\mathfrak{l},\mathfrak{p}]\subseteq \mathfrak{p},
[\mathfrak{p},\mathfrak{p}]\subseteq \mathfrak{l},
\end{eqnarray}
where we said the decomposition is the Cartan decomposition of Lie
algebra $\mathfrak{g}$. The $\mathfrak{l}$ is closed under the Lie
bracket, so it is a Lie subalgebra of $\mathfrak{g}$, and
$\mathfrak{p}=\mathfrak{l}^{\bot}$. A maximal Abelian subalgebra
$\mathfrak{a}$ contained in $\mathfrak{p}$ is called a Cartan
subalgebra. Then the element $M$ of Lie group $G$ can be decomposed
as
\begin{eqnarray}                                              \label{A3}
M=K_{1}AK_{2},
\end{eqnarray}
where  $G=e^\mathfrak{g}$, $K_{1}, K_{2}\in e^\mathfrak{l}$, and
$A\in e^\mathfrak{a}$.

For the qutrit case, we have eight independent ternary Pauli's
matrices: three $\sigma_{x}^{(ij)}$ matrices, three
$\sigma_{y}^{(ij)}$ matrices, and  two independent
$\sigma_{z}^{(ij)}$ matrices in the three of them. Multiplying these
eight  Pauli's matrices by $i$, we get the basis vectors of Lie
algebra $\mathrm{su}(3)$ which we called the qusi--spin basis.
Together with the $3\times3$ identity matrix multiplied by $i$, they
constitute the basis vectors of Lie algebra $\mathrm{u}(3)$. Take a
kind of AIII-type Cartan decomposition \cite{24} of $\mathrm{u}(3)$,
that is
\begin{eqnarray}            \label{A4}
\mathrm{u}(3)=\mathrm{s}(\mathrm{u}(2) \oplus
\mathrm{u}(1))\oplus \mathrm{s}(\mathrm{u}(2) \oplus
\mathrm{u}(1))^{\bot}.
\end{eqnarray}
 Lie subalgebra
$\mathrm{s}(\mathrm{u}(2) \oplus \mathrm{u}(1))$ consists of
subagebra $\mathrm{su}(2)$ and a complex basis
$r=diag\{I_2,-2\}=2\sigma_z^{(02)}-\sigma_z^{(01)}$. We choose
\begin{eqnarray}             \label{A5}
\mathrm{s}(\mathrm{u}(2) \oplus
\mathrm{u}(1))=\text{span}\{i(\sigma_x^{(01)},\sigma_y^{(01)},\sigma_z^{(01)},r)\}
\end{eqnarray}
and its Cartan subalgebra
\begin{eqnarray}                \label{A6}
\mathfrak{a}=\text{span}\{i(I_3,i\sigma_{y}^{(02)}\}.
\end{eqnarray}
So the one-qutrit matrix can be decomposed as
\begin{eqnarray}              \label{A7}
M&=&e^{i
\alpha}\tilde{M}_1^{(01)}R_z^{(01)}(-\theta)R_z^{(02)}(2\theta)R_y^{(02)}(\beta)\nonumber\\&&R_z^{(02)}(2\theta')
R_z^{(01)}(-\theta')\tilde{M}_2^{(01)}\nonumber\\
&=&e^{i \alpha}M_1^{(01)}M^{(02)}M_2^{(01)}.
\end{eqnarray}
Lie subalgebra and Cartan subalgebra of the Cartan decomposition can
be different, so the decomposition is not unique, and we can get the
more generic Eq. (\ref{eq1}) in Sec. \ref{sec2}.

For the generic qudit case, we can also use the qusi--spin basis.
There are  $\frac{1}{2}d(d-1)$ $\sigma_x^{(ij)}$ matrices,
$\frac{1}{2}d(d-1)$ $\sigma_y^{(ij)}$ matrices, and $d-1$
independent $\sigma_z^{(ij)}$ matrices for an $n$--dimensional
Hilbert space. Multiplying these $d^2-1$ independent qusi-spin
matrices by $i$, we gain the basis vectors of the Lie algebra
$\mathrm{su}(d)$. Together with a $d\times d$ identity matrix
multiplied by $i$, they constitute the basis vectors of Lie algebra
$\mathrm{u}(d)$. We also take a kind of AIII-type Cartan
decomposition for $\mathrm{u}(d)$, that is,
\begin{equation}                                            \label{A8}
\mathrm{u}(d)=\mathrm{s}(\mathrm{u}(d-1)\oplus
\mathrm{u}(1))+\mathrm{s}(\mathrm{u}(d-1)\oplus u(1))^\perp.
\end{equation}
Lie algebra $\mathrm{s}(\mathrm{u}(d-1)\oplus \mathrm{u}(1))$
consists of subalgebra $\mathrm{su}(d-1)$ and a complex basis
$r=diag\{I_{d-1},-(d-1)\}$. We choose its Cartan subalgebra
\begin{eqnarray}                                        \label{A9}
\mathfrak{\alpha}=span\{i(I_d, \sigma_{y}^{(d-2,d-1)})\}.
\end{eqnarray}
So the arbitrary one--qudit matrix can be expressed as
\begin{eqnarray}                        \label{A10}
M=e^{i \alpha}K_{1}R_y^{(d-2,d-1)}(\beta)K_2,
\end{eqnarray}
where $K_i\in \mathrm{S}(\mathrm{U}(d-1)\oplus\mathrm{U}(1))$ group.
The matrix $M$ can be re-expressed as
\begin{eqnarray}                                             \label{A11}
M&=&e^{i \mathfrak{\alpha}}\tilde{K'_1}e^{i\theta
r}R_y^{(d-2,d-1)}(\beta)e^{i\theta'r} \tilde{K'_2}\nonumber\\
&=&e^{i \mathfrak{\alpha}}K'_{1}M^{(d-2,d-1)} K'_2.
\end{eqnarray}
where $\tilde{K'_i},K'_i\in \mathrm{SU}(d-1)\oplus 1$. That is
because that $r$ can be expressed as a linear combination of
$\sigma_z^{(jk)}$s,
$r=\sigma_z^{(0,\,d-2)}+\cdots+\sigma_z^{(d-3,d-2)}+(d-1)\sigma_z^{(d-2,\,d-1)}$,
so  the  $e^{i\theta r}$ is a product of a series of $R_z^{(jk)}$s.
The $R_y^{(d-2,\,d-1)}$ combines with $R_z^{(d-2,\,d-1)}$ in
$e^{i\theta r}$ and $e^{i\theta r'}$ to form the $M^{(d-2,\,d-1)}$;
other $R_z^{(jk)}$s are absorbed in  $K_i'$s.

From Eq. (\ref{A11}) we can see that the $d$-dimensional one-qudit
elementary gates need one pair of $R_\alpha^{(jk)}$ gates more than
that for the ($d-1$)-dimensional qudit. They come from Euler
decomposition of $M^{(d-2,\,d-1)}$. The ($d-1$)-dimensional qudit
matrix $K'$ can be decomposed further in the same mode. The
successive decomposition can be done until the qutrit occurs. So we
can infer that the set of $d$-dimensional one-qudit elementary gates
has $d-1$ pairs of $R_\alpha^{(jk)}$ gates.

\section{SYNTHESIS OF $\pmb{p\wedge_2(X)}$ GATE} \label{B}

Many syntheses of gates given in Sec. \ref{sec4} can be verified by
matrix computing. In the simplest case of Fig. \ref{Fig9}, we get
the $p\wedge_2(X)$ gate. For $d=3$, $m=2$, $m'=2$,
$X^{(ij)}=X^{(12)}$, we calculate the matrix
\begin{eqnarray}                       \label{B1}
M&=&(I_3\otimes I_3 \otimes R_y^{(12)}(-\pi/4))\cdot(I_3 \otimes
\text{\text{\scriptsize{GCX}}})\nonumber\\&&\cdot (I_3\otimes I_3
\otimes
R_y^{(12)}(-\pi/4))\cdot\text{\scriptsize{GCX}}(1\rightarrow3)
\nonumber\\&&\cdot (I_3\otimes I_3 \otimes
R_y^{(12)}(\pi/4))\cdot(I_3 \otimes \text{\scriptsize{GCX}})
\nonumber\\&&\cdot(I_3\otimes I_3 \otimes R_y^{(12)}(\pi/4)).
\end{eqnarray}
The result is
\begin{eqnarray}                       \label{B2}
M=diag\{I_{18},Z^{(12)},Z^{(12)},X^{(12)}\},
\end{eqnarray}
where
\begin{eqnarray}                       \label{B3}
Z^{(12)}=\left(\begin{array}{ccc}
1&0&0\\
0&1&0\\
0&0&-1\\
\end{array}\right),\;\;
X^{(12)}=\left(\begin{array}{ccc}
1&0&0\\
0&0&1\\
0&1&0\\
\end{array}\right).
\end{eqnarray}
If we calculate
\begin{eqnarray}                       \label{B4}
M'&=&(I_3\otimes I_3 \otimes R_y^{(12)}(\pi/4))\cdot(I_3 \otimes
\text{\scriptsize{GCX}})\nonumber\\&&\cdot (I_3\otimes I_3 \otimes
R_y^{(12)}(\pi/4))\cdot \text{\scriptsize{GCX}}(1\rightarrow3)
\nonumber\\&&\cdot (I_3\otimes I_3 \otimes
R_y^{(12)}(-\pi/4))\cdot(I_3 \otimes \text{\scriptsize{GCX}})
\nonumber\\&&\cdot(I_3\otimes I_3 \otimes R_y^{(12)}(-\pi/4)),
\end{eqnarray}
the result is
\begin{eqnarray}                       \label{B5}
M'=diag\{I_{18},Z^{(21)},Z^{(21)},X^{(12)}\},
\end{eqnarray}
where
\begin{eqnarray}                       \label{B6}
Z^{(21)}=\left(\begin{array}{ccc}
1&0&0\\
0&-1&0\\
0&0&1\\
\end{array}\right). \end{eqnarray} The $M$ and $M'$ satisfy the
definition of $p\wedge_2(X)$ gate. Likewise, the syntheses of the
$\wedge_2(\text{\scriptsize{INC}})$ gates, the generic
$p\wedge_k(X)$ gates and so on have been verified.

\section{SYNTHESIS OF MULTI-VALUED UNIFORMLY MULTI-FOLD CONTROLLED $\pmb{R^{(ij)}_\alpha} \pmb{(\alpha \in y,z)}$  ROTATION} \label{C}

\begin{figure}[!htb] \begin{center} \includegraphics[width=8.5
cm,angle=0]{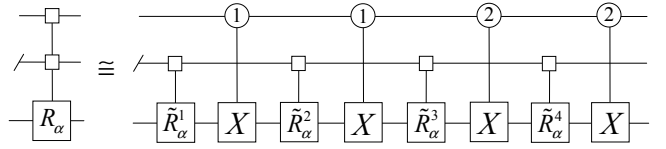} \caption{First step decomposition of a
uniformly multifold controlled $R^{(ij)}_\alpha$  rotation.}
\label{FigA} \end{center} \end{figure}

Taking $d=3$  as an example, the first step of the decomposition is
shown in Fig. \ref{FigA}. It involves four \text{\scriptsize{GCX}}
gates and four uniformly $(k-1)$-fold controlled $R^{(ij)}_\alpha$
rotations.

The second step is to decomposition the four uniformly  $(k-1)$-fold
controlled rotations. It produces eight \text{\scriptsize{GCX}}
gates and 12 uniformly $(k-2)$-fold controlled rotations. In the
process, four pairs of \text{\scriptsize{GCX}} gate cancel, and four
pairs of uniformly $(k-2)$-fold controlled rotation are combined.
The uniformly $(k-2)$-fold controlled rotation can be decoupled
further. The method can be used to generic case. The first step
produces a $2(d-1)$ \text{\scriptsize{GCX}} gate, the second step
produces $2(d-1)^2$ \text{\scriptsize{GCX}} gates, and so on.
Totally, it needs $2d^{k-1}(d-1)$ \text{\scriptsize{GCX}} gates.

The quantum circuit implementing ternary uniformly two-fold
controlled $R_\alpha$ rotation is shown in Fig. \ref{Fig12}. It has
also been verified by matrix computing.


\end{document}